%Paper: hep-th/9402127
%From: fteorica@cpd.uva.es
%Date: Tue, 22 FEB 94 22:14 GMT

%%%%%%%%%%%%%%%%%%%%% AMS-TEX FILE %%%%%%%%%%%%%%%%%%%%%

\magnification\magstep1
\hoffset=4pt
\parindent=1.5em
\TagsOnRight
\baselineskip=12pt
\NoBlackBoxes
\font\ninepoint=cmr9
\NoBlackBoxes

            % los n
meros reales
            % los n
meros enteros

\define\>#1{{\bold#1}}                 %  notaci
n para vectores

\define\co{\Delta}                     %  coproducto
\define\conm#1#2{\left[ #1,#2 \right]} %  un conmutador
\def\1{\'{\i}}                         %  i con acento

\def\Dr{1}
\def\FRT{2}
\def\Bu{3}
\def\CGH{4}
\def\CGST{5}
\def\BGST{6}
\def\HLR{7}
\def\Kup{8}
\def\CGP{9}

\font\cabeza=cmbx12

\centerline{\cabeza A UNIVERSAL NON QUASITRIANGULAR QUANTIZATION}
\smallskip
\centerline{\cabeza OF THE HEISENBERG GROUP}
\bigskip \smallskip

\centerline {A. Ballesteros$^{1),3)}$, E. Celeghini$^{2)}$, F.J.
Herranz$^{1)}$, M.A. del Olmo$^{1)}$ and M. Santander$^{1)}$}
\bigskip
\centerline{\it $^{1)}$Departamento de F\1sica Te\'orica, Universidad
de Valladolid.}
\centerline{\it E-47011 Valladolid, Spain.}
\smallskip
\centerline{\it $^{2)}$Dipartimento di Fisica and INFN, Sezione di
Firenze.}
\centerline{\it Largo E. Fermi, 2. 50125-Firenze, Italy.}
\smallskip
\centerline{\it $^{3)}$Departamento de F\1sica Aplicada III, E.U.
Polit\'ecnica.}
\centerline{\it E-09006 Burgos, Spain.}

\smallskip

\bigskip

\ninepoint
\noindent ABSTRACT. A universal $R$--matrix
for the quantum Heisenberg algebra $\frak h(1)_q$ is presented.
Despite of the non--quasitriangularity of this Hopf algebra, the
quantum group induced from it coincides with the quasitriangular
deformation already known.
\medskip

%\noindent KEYWORDS: Quantum algebras, space--time symmetries.

\tenpoint
\bigskip\medskip
\bigskip\medskip

Recall that a quasitriangular Hopf algebra [\Dr] is a pair $(\Cal
A,\Cal R)$ where $\Cal A$ is a Hopf algebra and $\Cal R \in \Cal
A\otimes \Cal A$ is invertible and verifies
$$
\align
\sigma \circ \co h = \Cal R (\co h) \Cal R^{-1}, &\qquad \forall\
\, h \in \Cal A \tag 1.a
\\ (\co \otimes id)\Cal R =\Cal R_{13}\Cal
R_{23},\qquad & (id \otimes \co)\Cal R =\Cal R_{13}\Cal R_{12}, \tag
1.b \endalign
$$
where, if $\Cal R=\sum_{i}
a_i\otimes b_i$,  we denote $\Cal R_{12}\equiv\sum_{i} a_i\otimes
b_i\otimes 1$, $\Cal R_{13}\equiv\sum_{i} a_i\otimes 1\otimes b_i$
,$\Cal R_{23}\equiv\sum_{i} 1\otimes a_i\otimes b_i$ and $\sigma$
is the flip operator $\sigma(x\otimes y)=(y\otimes x)$. If $\Cal A$
is a quasitriangular Hopf algebra, then  $\Cal R$ is called an
``universal"  $\Cal R$--matrix and satisfies the Quantum
Yang--Baxter Equation (QYBE):
$$
\Cal R_{12}\Cal R_{13}\Cal R_{23}=\Cal
R_{23}\Cal R_{13}\Cal R_{12}.
\tag 2
$$

Given a matrix representation $\rho:\Cal A\rightarrow
\text{Mat}(n,\Bbb C)$, the FRT [\FRT] approach to quantum groups is
based on an $R$-matrix
taken as $R=(\rho\otimes\rho)(\Cal R)$. The matrix entries $t_{ij}
(i,j=1,\dots,n)$ of a general ``quantum group" element $T$ satisfy
the commutation relations $$
R \, T_1 \, T_2 \,= \, T_2 \, T_1 \, R, \tag 3
$$
where $T=(t_{ij})$, $T_1=T\otimes 1_n$ and $T_2=1_n\otimes T$. The
Yang--Baxter condition ensures the associativity of the
(non--commutative) algebra of functions on the quantum group
$Fun(G_q)$ generated by the $t_{ij}$. It also provides a
straightforward way to obtain $\Cal A$ as the dual of $Fun(G_q)$ by
using the $L^\pm$ matrices [\Dr,\Bu].

\medskip

It would be worth to study whether the same $Fun(G_q)$ can be
derived with the aid of the FRT prescription and using a non
quasitriangular $R$-matrix which solves (1.a). This letter shows
that, if we consider the quantum Heisenberg algebra $\frak h(1)_q$
[\CGH,\CGST], such a universal $R$-matrix can be found. Moreover,
its associated quantum Heisenberg group coincides with the one given
in [\CGH,\BGST] (see [\HLR] for a characterization of all possible
quantizations of this group).

\medskip

Let us consider an $R$-matrix as a formal
power series in $z$ with coefficients in $U \frak h(1) \otimes
U \frak h(1)$:
$$
R=e^{zR_1} + \sum_{i=1}^\infty z^i (e^{zR_{i+1}}-1).\tag 4
$$
Provided the coproduct $\co$ is also expanded in terms of $z$, we
can try to solve (1.a) order by order in the deformation parameter
and obtain the explicit form of the $R_i$ components.

\medskip

The quantum Heisenberg algebra $\frak h(1)_q$
[\CGH] is defined by the commutation relations
$$
\conm{A}{A^\dagger}={{\sinh z H}\over{z}}, \quad
\conm{A}{H}=0, \quad
\conm{A^\dagger}{H}= 0,
\tag 5
$$
which are consistent with the Hopf homomorphisms
$$
\aligned
\co(H)&=1\otimes H + H\otimes 1,\\
\co(A^\dagger)&=e^{-{z\over 2}
H}\otimes A^\dagger + A^\dagger\otimes e^{{z\over2}H},\\
\co(A)&=e^{-{z\over 2} H}\otimes A + A\otimes
e^{{z\over 2} H};
\endaligned \tag 6
$$
$$
\epsilon(X) =0;\qquad
\gamma(X)=-e^{{z\over 2}  H}\ X\ e^{-{z\over 2}  H}=-X, \qquad
X\in\{ A, A^\dagger, H\}.
\tag 7
$$

If we assume the Ansatz (4) for the $R$-matrix, the
first order of (1.a) leads to
$$
\conm{R_1}{1\otimes X + X\otimes 1}=H\otimes X - X\otimes H,
\tag 8
$$
whose solution is
$$
R_1=A\otimes A^\dagger - A^\dagger\otimes A.
\tag 9
$$
Note that, as expected, $R_1$ is a classical $r$--matrix
for the algebra $\frak h(1)$ (it fulfills the modified classical
YB equation). The study of the $r$--matrices for this algebra has
been given in [\BGST], where it is shown that (9) provides the
(essentially unique) Poisson--Lie quantization of $H(1)$.

\medskip

The second order leads to the condition
$$
\conm{R_2}{1\otimes X + X\otimes 1}=0 \tag 10
$$
for any generator $X$. This requirement is obviously fulfilled if
$R_2=0$, and we shall adopt this solution.

\medskip

The third order originates the following equation for $R_3$
$$
\aligned
\conm{R_3}{1\otimes X+X\otimes 1}&=\tfrac
1{12}\conm{R_1}{\conm{R_1}{\conm{R_1}{1\otimes X+X\otimes 1}}} -
\tfrac 1 8 \conm{R_1}{H^2\otimes X+X\otimes H^2} \\
&\qquad -\tfrac 1 8 \conm{R_1}{H^3\otimes X+X\otimes H^3}.
\endaligned
\tag 11
$$
A solution for (11) is given by
$$
R_3=-R_1\left(\tfrac 1 {12} H\otimes H + \tfrac 1 8 (1\otimes
H^2 +H^2\otimes 1) \right). \tag 12
$$

Note that $R_3$ exhibits a remarkable property: it depends only on
$R_1$ and $H$. We put forward this dependence by supposing
that $R$ can be written as
$$
R=e^{z R_1 f(H,z)}. \tag 13
$$
Under this assumption, the condition (1.a) can be solved in
general (from now on, we shall not write the $z$-dependence of the
function $f$). Explicitly,
$$
\aligned
R \co X (R)^{-1}&=e^{z R_1 f(H)} \co X e^{-z R_1 f(H)} \\
&=\co X + z f(H) \conm{R_1}{\co X} + \tfrac 1 {2!} z^2  f(H)^2
\conm{R_1}{\conm{R_1}{\co X}} + \dots \\
&\qquad + \tfrac 1 {n!} z^n  f(H)^n
\conm{R_1}{\left[\dots\conm{R_1}{\co X}\right]^{n)}\dots} + \dots
\endaligned \tag 14
$$
For $X\equiv A$ and $X\equiv A^\dagger$ we need to obtain the
brackets in (14):
$$
\align
\conm{R_1}{\co X}&=
\conm{A\otimes A^\dagger - A^\dagger\otimes A}{e^{-{z\over 2}
H}\otimes X + X\otimes e^{{z\over2}H}}\\
&=\tfrac 1 z \sinh (z H)\otimes X e^{{z\over 2}H}
- X e^{-{z\over2}H}\otimes \tfrac 1 z \sinh (z H), \tag 15 \\
\conm{R_1}{\conm{R_1}{\co X}}&=
-\tfrac 1 {z^2} (\sinh z H\otimes \sinh z H)\co X, \tag 16 \\
\conm{R_1}{\conm{R_1}{\conm{R_1}{\co X}}}&= -\tfrac 1 {z^2}
(\sinh z H\otimes \sinh z H)\conm{R_1}{\co X}. \tag 17
\endalign
$$
In general, a recurrence method gives
$$
\align
\conm{R_1}{\left[\dots\conm{R_1}{\co
X}\right]^{2n)}\dots}&={{(-1)^n}\over{z^{2n}}}(\sinh z H \otimes
\sinh z H)^n \co X, \tag 18 \\
\conm{R_1}{\left[\dots\conm{R_1}{\co
X}\right]^{2n + 1)}\dots}&={{(-1)^n}\over{z^{2n+1}}}(\sinh z H
\otimes \sinh z H)^n z \conm{R_1}{\co X}. \tag 19
\endalign
$$
Now, expression (14) can be written as follows:
$$
\align
e^{z R_1 f(H)} \co X e^{-z R_1 f(H)}&=\sum_{l=0}^\infty
{{(-1)^l}\over{(2l)!}} f(H)^{2l} (\sinh z H \otimes
\sinh z H)^l \co X \\
&\qquad + \sum_{j=0}^\infty
{{(-1)^j}\over{(2j+1)!}} f(H)^{2j+1} (\sinh z H \otimes
\sinh z H)^j z \conm{R_1}{\co X} \\
&=\cos\left(\sqrt{\sinh z H \otimes \sinh z H} f(H) \right) \co X \\
&\qquad
+ {{\sin\left(\sqrt{\sinh z H \otimes \sinh z H} f(H)
\right)}\over{\sqrt{\sinh z H \otimes \sinh z H}}} z \conm{R_1}{\co
X} \tag 20.a\\
&=\alpha(H) \co X + \beta(H) z \conm{R_1}{\co X}, \tag 20.b
\endalign
$$
where $\alpha(H)$ and $\beta(H)$ are to be determined. We recall
that (20) has to provide the same effect on $\co X$ as the flip
operator $\sigma$. Explicitly, for $X\neq H$ we have
$$
\align
\sigma \circ \co X &= e^{{z\over 2} H}\otimes X +
X\otimes e^{-{z\over 2} H}\\
&=(X\otimes 1)(1 \otimes e^{-{z\over 2} H}) + (e^{{z\over 2}
H}\otimes 1)(1 \otimes X). \tag 21
\endalign
$$
If we impose (21) to coincide with (20.b), we obtain the
following system of equations for $\alpha$ and $\beta$:
$$
\aligned
\alpha (H) (e^{-{z\over 2}H}\otimes 1) + \beta (H)(\sinh zH \otimes
e^{{z\over 2}H})&=(e^{{z\over 2}H}\otimes 1)\\
\alpha (H) (1\otimes e^{{z\over 2}H}) - \beta (H)(e^{-{z\over
2}H}\otimes \sinh zH )&=(1\otimes e^{-{z\over 2}H}).
\endaligned
\tag 22
$$
These equations can be solved and lead to the solution
$$
\aligned
\alpha(H)&={{1\otimes 1 + e^{-zH}\otimes e^{zH}}\over{1\otimes
e^{zH} + e^{-zH}\otimes 1}}=2 {{e^{{z\over 2}H}\otimes
e^{-{z\over 2}H} + e^{-{z\over 2}H}\otimes e^{{z\over
2}H}}\over{\cosh(\tfrac z 2 \co H)}},\\
\beta(H)&={{2}\over{e^{{z\over 2}H}\otimes
e^{{z\over 2}H} + e^{-{z\over 2}H}\otimes e^{-{z\over
2}H}}}={{1}\over{\cosh(\tfrac z 2 \co H)}}.
\endaligned
\tag 23
$$
For the sake of consistency between (20) and (23), it can be
easily checked that $\alpha(H)^2 + \sinh (zH) \otimes \sinh (zH)
\beta(H)^2=1$. Finally, from these two equations we deduce the final
form of the function $f(H)$ which reads
$$
\align
f(H,z)&={{1}\over{\sqrt{\sinh z H \otimes \sinh z H}}} \arcsin \left(

{{\sqrt{\sinh z H \otimes \sinh z H}}\over{\cosh(\tfrac z 2 \co
H)}} \right),\quad \text{or} \tag 24 \\
f(H,z)&={{1}\over{\sqrt{\sinh z H \otimes \sinh z
H}}} \arccos \left(  2 {{e^{{z\over 2}H}\otimes
e^{-{z\over 2}H} + e^{-{z\over 2}H}\otimes e^{{z\over
2}H}}\over{\cosh(\tfrac z 2 \co H)}} \right). \tag 25
\endalign
$$
A power series expansion of $f(H,z)$ in terms of $z$ leads to (4),
with $R_1$ and $R_3$ given by (9) and (12), respectively. The
$R$--matrix (13) so obtained does not fulfill QYBE equation.

\medskip

Let us consider now a three dimensional representation of
the Heisenberg group
$$
T=\pmatrix 1 & \theta & a_1 \\
0 & 1 & a_2 \\
0 & 0 & 1 \endpmatrix. \tag 26
$$

The corresponding quantum group is the (non--commutative) Hopf
algebra of functions generated by the matrix
entries $t_{ij}$. Their coproduct is induced from the group
multiplication $T\dot\otimes T$:
$$
\aligned
\co (\theta)&=1\otimes \theta + \theta\otimes 1,\\
\co(a_1)&=1\otimes a_1  + a_1\otimes 1 + \theta\otimes a_2,\\
\co(a_2)&= 1\otimes a_2 + a_2\otimes 1.
\endaligned
\tag 27
$$
Counit and antipode are derived from the unit matrix and from
$T^{-1}$, respectively (see [\CGH]).

\medskip

The fundamental representation of the
quantum algebra (5) is
$$
D(H)=\pmatrix
0&0&1\\0&0&0\\0&0&0\endpmatrix ,\quad D(A^\dagger)=\pmatrix
0&0&0\\0&0&1\\0&0&0\endpmatrix ,\quad  D(A)=\pmatrix
0&1&0\\0&0&0\\0&0&0\endpmatrix.\tag 28
$$
If we realize the $R$--matrix (13) in terms of (28) we find that
only the linear term in $z$ is left (see also [\Kup]):
$$
\align
D(R)&=1_3 + z \left[ D(A)\otimes D(A^\dagger) - D(A^\dagger)\otimes
D(A) \right] \\
&=\pmatrix 1_3 & z D(A^\dagger) &0\\
           0&1_3&-z D(A)\\
           0&0&1_3 \endpmatrix. \tag 29
\endalign
$$

This $R$--matrix, together with (3) and (26), provides
$$
\aligned
\conm{\theta}{a_1}&= z \theta\\
\conm{\theta}{a_2}&= 0 \\
\conm{a_1}{a_2}&= - z a_2,
\endaligned
\tag 30
$$
This algebra verifies Jacobi
identity and coincides (up to a change in the deformation parameter
$z=w/2$) with the quantum Heisenberg group obtained from a
quasitriangular Hopf algebra in [\CGH]. In that paper, a
number operator $N$ was needed in order to obtain a universal $\Cal
R$--matrix, in the same way as its classical counterpart $n$ has been
introduced to generate the same $\Cal R$--matrix from the coboundary
Poisson--Lie structure studied in [\BGST]. On the other hand, the
duality between relations (27,30) and the quantum Heisenberg
algebra (5) has been given in [\CGP]. Finally, note that the
universality of $R$ allows the FRT construction for any finite
dimensional  representation of the Heisenberg group.

\bigskip
\bigskip

\noindent {\bf Acknowledgments}
\medskip

This work has been partially supported by a DGICYT project
(PB92--0255) from the Ministerio de Educaci\'on y Ciencia de
Espa\~na and by an Acci\'on Integrada Hispano--Italiana (HI--059).

\bigskip
\bigskip

\noindent\bf References
\rm
\medskip
\eightpoint

\ref      %% 1
\no[{\Dr}]
\by V.\, G.\, Drinfeld
%\paper Quantum Groups
\jour Proceedings of the International Congress of Mathematics,
MRSI Berkeley (1986), 798
\endref

\ref     %% 2
\no[{\FRT}]
\by      N. Yu. Reshetikhin, L.A. Takhtadzhyan and L.D. Faddeev
\jour    Leningrad Math. J.
\vol     1
\yr      1990
\pages   193
\endref

\ref     %% 3
\no[{\Bu}]
\by      N. Burroughs
\jour    Comm. Math. Phys.
\vol     133
\pages   91
\yr      1990
\endref

\ref     %% 4
\no[{\CGH}]
\by      E. Celeghini, R. Giachetti, E. Sorace, and M. Tarlini
\jour    J. Math. Phys.
\vol     32
\pages   1155
\yr      1991
\endref

\ref      %% 5
\no[{\CGST}]
\by E. Celeghini, R. Giachetti, E. Sorace and M. Tarlini
\paper Contractions of quantum groups
\jour Lecture Notes in Mathematics, n. 1510, Springer--Verlag
\pages 221
\endref

\ref     %% 6
\no[{\BGST}]
\by      F. Bonechi, R. Giachetti, E. Sorace, and M. Tarlini
\paper   Deformation Quantization of the Heisenberg Group
\jour    DFF preprint,
\yr      1993
\endref

\ref     %% 7
\no[{\HLR}]
\by      V. Hussin, A. Lauzon and G. Rideau
\paper   $R$-matrix method for Heisenberg quantum groups
\jour    preprint CRM--1917, Montr\'eal
\yr      1994
\endref

\ref     %% 8
\no[{\Kup}]
\by      B.A. Kuperschmidt
\jour    J. Phys. A: Math. Gen
\vol     26
\pages   L929
\yr      1993
\endref

\ref     %% 9
\no[{\CGP}]
\by      F. Bonechi, E. Celeghini, R. Giachetti, C.M. Pere\~na, E.
Sorace, and M. Tarlini
\paper   Exponential Mapping for Non-semisimple Quantum Groups
\jour    To be published in J. Phys. A: Math. Gen
%\yr      1993
\endref

\end